# The ideas behind the Self Consistent Expansion

Moshe Schwartz

School of Physics and Astronomy, Raymond and Beverly Sackler

Faculty of Exact Sciences,

Tel-Aviv University, Tel-Aviv 69978, Israel

and

Eytan Katzav

Laboratoire de Physique Statistique, Ecole Normale Supérieure,

24 rue Lhomond, 75231 Paris Cedex 05, France

abstract

In recent years we have witnessed a growing interest in various non-equilibrium systems described in terms of stochastic non-linear field theories. In some of those systems like the KPZ and related models, the interesting behavior is in the strong coupling regime, which is inaccessible by traditional perturbative treatments such as dynamical renormalization group (DRG). A useful tool in the study of such system is the Self Consistent Expansion (SCE), which might be said to generate its own "small parameter" .The self consistent expansion (SCE) has the advantage that its structure is just that of a regular expansion, the only difference is that the simple system around which the expansion is performed is adjustable. The purpose of the this article is to present the method in a simple and understandable way, that hopefully will make it accessible to a wider public working on non-equilibrium statistical physics.

**Introduction**

The focus of interest in the study of dynamical statistical problems has shifted in the last two decades from equilibrium phase transitions and later the dynamics of phase transitions [1] to the study of non-equilibrium systems which is far richer and many intriguing scaling phenomena, such as self-organized criticality [2], or phase transitions between non-equilibrium stationary states [3], have been observed for long. The list of systems intensively studied includes various growth models [4,5,6,7], front propagations [6,8,9] , crack propagation [10,11] etc. In spite of that shift, the main objects of study remained of a similar nature, a small set of exponents which describe the steady state properties as well as the evolution of the system.

Naturally the renormalization group (RG), proven useful to explain universality in equilibrium continuous phase transitions and to obtain the critical exponents was the obvious method to turn to. Indeed its use has allowed some progress in understanding some systems out-of-equilibrium. Nevertheless, in many cases the information RG analysis offers is inherently not complete. A classical example is the Kardar-Parisi-Zhang (KPZ) equation [4] where the Dynamic Renormalization Group (DRG) approach has no access to the properties of the strong coupling phase apart from the case of one dimension where it agrees with the analytic exact result [6]. In addition, a remarkable result of Wiese [12] shows that the shortcoming of DRG in the KPZ system is not an artifact of a low order calculation, but rather intrinsic to the method and extends to all orders. Indeed the DRG results for KPZ, obtained while ignoring the above, are in total disagreement with reliable simulations [6]. Even in one dimension, when derivatives of the original KPZ problem have been studied, the DRG failed [7] to produce the exact results [13].

The great advantage of DRG is that it is based on a perturbation expansion – a technique with which the whole community of theoretical physicist is familiar. Its weakness is that in general it is hard to expect a weak coupling expansion to produce a strong coupling result. An alternative approach which technically is also a perturbation expansion is the Self Consistent Expansion introduced by Schwartz and Edwards to study the KPZ problem [14]. The main idea is that an expansion should always be optimized by choosing the zero system in such a way that it already mimics the full system under consideration. The choice is done a posteriori rather than a priori. This will become very clear in the zero-dimensional problem we discuss following the introduction.

The SCE was first applied to the KPZ equation [14]. In one dimension it produces the exact result and in more than one dimension it produces exponents in agreement to those obtained by simulations. In more than two dimensions it produces also the weak coupling solution (that is possible above two dimensions) in addition to the strong coupling solution [15]. The method has been successfully applied to KPZ systems with noise that is algebraically correlated in space [16] and in time [17], to the MBE [18] equation and to a family of non-local models such as the Non-local KPZ equation [19]. For the exactly soluble families of one dimensional variants of KPZ, where DRG fails, SCE produces all the exact results. SCE predicts stretched exponential decay of the KPZ time dependent structure factor [20], which was later verified in one dimension by numerical integration of the KPZ equation [21]. The method was also used for the study of vortex lines in the three dimensional X-Y model with random phase shifts [22], Turbulence [23] and recently to wetting and fracture [9,11]. In spite of its success the use of SCE is not widely spread and the main purpose of the present article is to make it accessible to a wider public by explaining its ideology and the way it is implemented. This will mostly be done by considering a zero-dimensional example, which can be solved numerically and allows therefore, assessing the strength the approximation. People speak some times about controlled vs. uncontrolled approximations. The regular expansion in the coupling strength is "controlled" of course, in spite of the fact that it may be a non-convergent (asymptotic) expansion. In the zero-dimensional results, the actual performance of the "controlled" approximation vs. that of the SCE, which is shown to converge, is presented.

The paper is organized as follows. In section A we present the SCE treatment of the anharmonic oscillator. In section B we construct the analogy for a KPZ type field theory by constructing the SCE for the two-point function and obtain an integral equation relating the steady state two-point function and the corresponding typical frequency. Next we describe the SCE for the typical frequency and obtain a second equation relating the two functions. We show in section C how to obtain an integral equation for the Fourier transform of the time dependent structure factor.

**A. SCE for a single degree of freedom-the anharmonic oscillator**

We discuss the anharmonic oscillator described by the following Langevin equation

$$\frac{\partial \phi(t)}{\partial t} = -\gamma \phi(t) - g\phi^3(t) + \eta(t), \tag{1}$$

where $\eta(t)$ is a Gaussian noise term satisfying

$$\langle \eta(t)\eta(t') \rangle = 2D_0 \delta(t-t'). \tag{2}$$

Quantities of interest are the static (equal-time) point correlation functions such as $\langle \phi^2 \rangle$, and more generally time dependent quantities such as the two-time two-point function $\langle \phi(0)\phi(t) \rangle$.

The Self-Consistent approach is typically (but not necessarily [23,17]) performed using the Fokker-Planck equation associated with the Langevin equation (1), here given by

$$\frac{\partial P}{\partial t} - \frac{\partial}{\partial \phi}\left( D_0 \frac{\partial}{\partial \phi} + \gamma\phi + g\phi^3 \right)P = 0, \tag{3}$$

where $P(\phi,t)$ is the probability distribution of the $\phi$'s at time $t$. It is easy to verify that a steady state exists and is described by

$$P \propto \exp\left[ -\frac{1}{D_0}\left( \frac{\gamma}{2}\phi^2 + \frac{g}{4}\phi^4 \right) \right]. \tag{4}$$

This property is directly related to the existence of a Hamiltonian for this system, as the right hand side of eq. (4) is just a Boltzmann factor. Thus, equal time quantities such as $\langle \phi^2 \rangle$ can be simply obtained from integrals such as

$$\left\langle \phi^2 \right\rangle = \frac{\int\limits_{-\infty}^{\infty} \phi^2 e^{-\frac{1}{D_0}\left(\frac{\gamma}{2}\phi^2 + \frac{g}{4}\phi^4\right)} d\phi}{\int\limits_{-\infty}^{\infty} e^{-\frac{1}{D_0}\left(\frac{\gamma}{2}\phi^2 + \frac{g}{4}\phi^4\right)} d\phi}. \qquad (5)$$

This boils down to determining the free energy of an anharmonic oscillator, from which everything follows. Still, note that for field theories, the simple integrals in eq. (5) are replaced by functional integrals, and become highly nontrivial.

We would like, however, to take another route, and derive results directly from the Fokker-Planck formulation for two reasons. First, it allows studying also dynamical properties such as the two-time quantity $\left\langle \phi(0)\phi(t) \right\rangle$. Second, as we are interested in nonlinear statistical field theories the existence of a Hamiltonian is not obvious, and relations such as (5) are the exception rather than the rule.

Before describing SCE, let us start with a warm up exercise and calculate the static two-point function $\left\langle \phi^2 \right\rangle$ as a power series in the nonlinear coupling $g$. This will help to gain acquaintance with eq. (3), and later on to assess the results of SCE.

First, since we are interested in the static quantity $\left\langle \phi^2 \right\rangle$, we drop the time derivative from eq. (3), multiply it by $\frac{1}{2}\phi^2$., and integrate with respect to $\phi$. After some integration by parts we get

$$D_0 - \gamma\left\langle \phi^2 \right\rangle - g\left\langle \phi^4 \right\rangle = 0. \qquad (6)$$

Next we multiply the equation for the steady state by $\frac{1}{4}\phi^4$ integrate by parts and obtain

$$3D_0\left\langle \phi^2 \right\rangle - \gamma\left\langle \phi^4 \right\rangle - g\left\langle \phi^6 \right\rangle = 0 \qquad (7)$$

etc.

This enables an easy expansion in $g$ that yields for the two-point function

$$\left\langle \phi^2 \right\rangle = \frac{D_0}{\gamma}\left[1 - 3\frac{D_0}{\gamma^2}g + 24\left(\frac{D_0}{\gamma^2}\right)^2 g^2 - 297\left(\frac{D_0}{\gamma^2}\right)^3 g^3 + \cdots\right]. \qquad (8)$$

Taking a look at the numerical pre-factors hints that this is a divergent series (or asymptotic). It is not difficult to show that it has zero radius of convergence since the point $g=0$ is special for the system described by eq. (1). In the absence of noise, for positive $g$'s $\phi$ approaches zero while for negative $g$'s it diverges. The effect of the noise is just to broaden the trajectory $\phi$ takes without it. Therefore, any expansion around $g=0$ should diverge. As can be seen in Fig. 1, the range of utility of this expansion is limited to very small values of g.

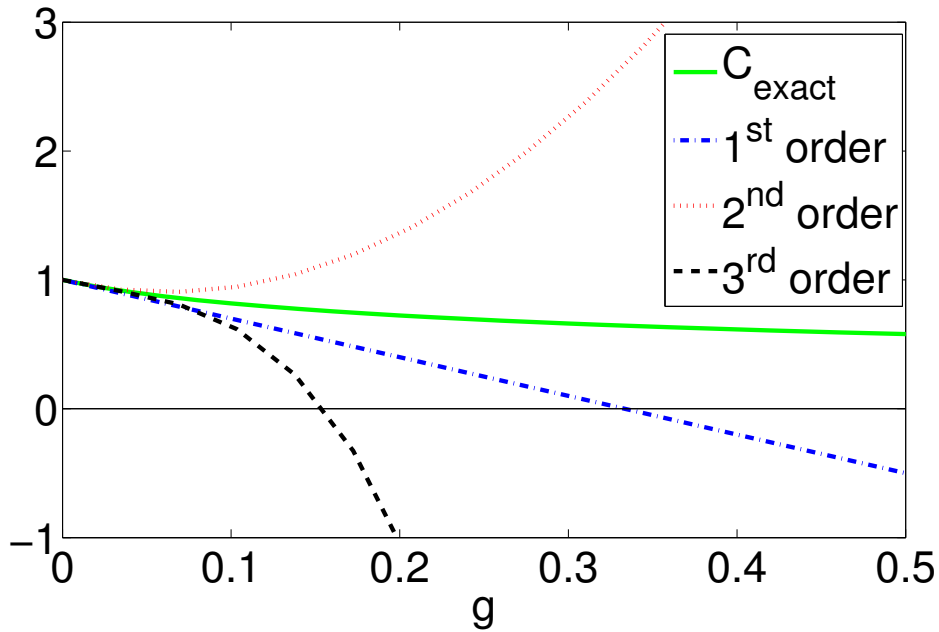

Fig. 1 – A comparison of the exact two-point function $\langle\phi^2\rangle$ with three successive power-series approximations of it, for $D_0=\gamma=1$. As can be seen the series deviates already for small values of $g$.

We now describe the Self Consistent Expansion. The main idea of SCE is to write the Fokker-Planck equation $\partial P/\partial t = OP$ in the form

$$\frac{\partial P}{\partial t} - \left[O^{(0)} + O^{(1)}\right]P = 0, \qquad (9)$$

where $O^{(0)}, O^{(1)}$ are zero/first order operators in some formal parameter $\Lambda$, which is used to keep track of the expansion. The evolution operator $O^{(0)}$ is chosen to have a simple form

$$O^{(0)} = \frac{\partial}{\partial \phi}\left(D_0 \frac{\partial}{\partial \phi} + \Gamma(g)\phi\right), \tag{10}$$

whose corresponding zeroth-order solution for $P^{(0)}$ is a Gaussian given by

$$P^{(0)} \propto \exp\left(-\frac{\Gamma(g)}{2D_0}\phi^2\right), \tag{11}$$

implying that at leading order the two-point function is given by $\langle \phi^2 \rangle^{(0)} = D_0/\Gamma(g)$. Note however that $\Gamma(g)$ is still not specified. Next, an equation for the two-point function is obtained. For that purpose, we first rewrite the Fokker-Planck equation (3) along the lines presented above

$$\frac{\partial P}{\partial t} - \frac{\partial}{\partial \phi}\left(D_0 \frac{\partial}{\partial \phi} + \Gamma(g)\phi + g\phi^3 + [\gamma - \Gamma(g)]\phi\right)P = 0, \tag{12}$$

To make our point we concentrate on the steady state case, $\partial P/\partial t = 0$. The full equation enables to obtain time dependent correlations such as $\langle \phi(0)\phi(t) \rangle$ but since our aim is to try and give a description which is as simple as possible we will not pursue it here.

An exact equation for the average of any quantity $\mathbb{F}(\phi)$, can be obtained by multiplying equation (3) by $\mathbb{F}(\phi)$ and integrating by parts. (A word of caution is in order here. The function $\mathbb{F}(\phi)$ must not diverge at $\pm \infty$ to strongly in order that the end point contribution when integrating by parts vanishes.)
We obtain

$$D_0\left\langle \frac{\partial^2 \mathbb{F}}{\partial \phi^2}\right\rangle - \gamma\left\langle \phi \frac{\partial \mathbb{F}}{\partial \phi}\right\rangle - \left\langle \phi^3 \frac{\partial \mathbb{F}}{\partial \phi}\right\rangle = 0. \tag{13}$$

The separation of the Fokker–Planck operator into a zero order part $O^{(0)}$, and a perturbation,

$$O^{(1)} = \frac{\partial}{\partial \phi}\left(g\phi^3 + [\gamma - \Gamma(g)]\right) \tag{14}$$

yields an equation relating the $n^{th}$ order approximants, $\langle ... \rangle^{(n)}$ to $(n-1)^{th}$ order approximants. Since, zero order approximants can be directly obtained, the iteration procedure presented by the next equation enables an expansion for the required averages as will be shown in the following.

$$D_0\left\langle\frac{\partial^2 \mathbb{F}}{\partial\phi^2}\right\rangle^{(n)} - \Gamma(g)\left\langle\phi\frac{\partial\mathbb{F}}{\partial\phi}\right\rangle^{(n)} - g\left\langle\phi^3\frac{\partial\mathbb{F}}{\partial\phi}\right\rangle^{(n-1)} - (\gamma-\Gamma(g))\left\langle\phi\frac{\partial\mathbb{F}}{\partial\phi}\right\rangle^{(n-1)} = 0, \quad (15)$$

Being interested in the two-point function, we insert $\mathbb{F} = \frac{1}{2}\phi^2$ into eq. (15). To zero order we obtain

$$D_0 - \Gamma(g)\left\langle\phi^2\right\rangle^{(0)} = 0, \quad (16)$$

from which follows $\left\langle\phi^2\right\rangle^{(0)} = D_0/\Gamma(g)$. Interestingly, this result can be easily generalized to any even moment (odd moments are trivially zero), where we get

$$\left\langle\phi^{2k}\right\rangle^{(0)} = \frac{(2k)!}{2^k k!}\frac{D_0}{\Gamma(g)^k}. \quad (17)$$

To first order, eq. (15) with $\mathbb{F} = \frac{1}{2}\phi^2$ becomes

$$D_0 - \Gamma(g)\left\langle\phi^2\right\rangle^{(1)} - g\left\langle\phi^4\right\rangle^{(0)} - (\gamma-\Gamma(g))\left\langle\phi^2\right\rangle^{(0)} = 0, \quad (18)$$

giving rise to

$$\left\langle\phi^2\right\rangle^{(1)} = \left[1 - \frac{\gamma-\Gamma(g)}{\Gamma(g)} - \frac{3g}{\Gamma(g)^2}\right]\frac{D_0}{\Gamma(g)}. \quad (19)$$

These results are of course meaningless unless one chooses an appropriate effective friction $\Gamma(g)$. Here comes the core of the self-consistent expansion, where we impose that the lowest order expansion for the two-point function should be exact, in the highest order calculated. Namely, we force the perturbative correction into being zero. Thus it is as if, at least, for the required quantity, we are expanding in a small quantity, that does not change the initial value. For example $\left\langle\phi^2\right\rangle^{(0)} = \left\langle\phi^2\right\rangle^{(1)}$ above, gives rise to

$$\Gamma_{2,1}(g) = \frac{\gamma+\sqrt{\gamma^2 + 12gD_0}}{2}, \quad (20)$$

where the subscript 2,1 means that this $\Gamma(g)$ was obtained by imposing the exactness of the 2$^{nd}$ moment to 1$^{st}$ order. In general one could have required that the 2k$^{th}$ moment is exact to p$^{th}$ order, i.e. $\left\langle\phi^{2k}\right\rangle^{(0)} = \left\langle\phi^{2k}\right\rangle^{(p)}$, which yields $\Gamma_{2k,p}(g)$. Since $\langle \phi^{2k} \rangle^{(1)}$ is relatively easy to obtain, we can have a closed form expression for $\Gamma_{2k,1}(g)$

$$\Gamma_{2k,1}(g) = \frac{\gamma + \sqrt{\gamma^2 + 4gD_0(k+2)}}{2}. \tag{21}$$

Last, whatever $\Gamma(g)$ is obtained, it can be used in the expansion of $\langle \phi^2 \rangle$ to any desired order, say order n, namely $\langle \phi^2 \rangle^{(n)}$. To summarize, this provided a three-parameter family of approximations to two-point function which we now denote

$$C_{2k,p}^n \equiv \langle \phi^2 \rangle^{(n)} \Big|_{\Gamma = \Gamma_{2k,p}(g)}, \tag{22}$$

so that the real problem here is to choose a good candidate from this huge family. The convergence properties of this family are still largely unexplored. Still we can make the following remarks. We begin with the lowest order approximation within this approach, i.e. $C_{2,1}^1$, which can be written explicitly using eqs. (19), (20),

$$C_{2,1}^1 = \frac{2D_0}{\gamma + \sqrt{\gamma^2 + 12gD_0}}. \tag{23}$$

It turns out that already $C_{2,1}^1$ gives a very good approximation for the two-point function over a large range of parameters (see Fig. 2) – much better than a power series in $g$ can provide. Actually, when looking at the relative difference between $C_{2,1}^1$ and the exact value (see Fig. 3), it becomes clear that the maximal deviation is when $g \to \infty$, where the deviation is slightly higher than 10%. Furthermore when considering the subfamily $C_{2n,1}^n$ with $n = 1, 2, 3, \cdots$ the approximation is monotonously improved for increasing $n$'s. As can be seen in Fig. 3, the relative errors falls below 4% already for $n = 3$.

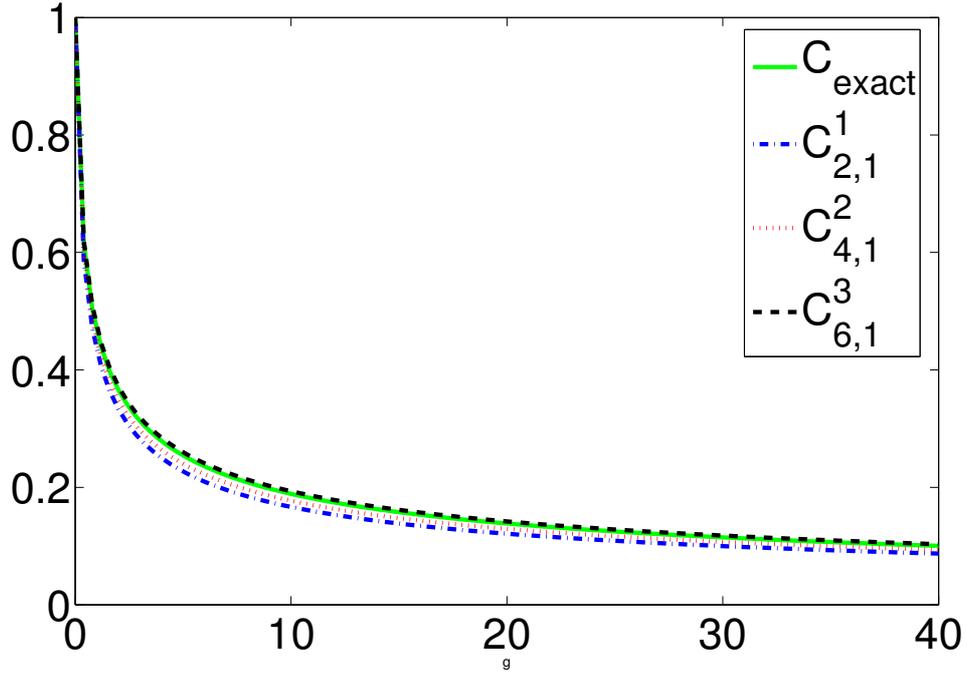

Fig. 2 – A comparison of the exact two-point function $\langle\phi^2\rangle$ with three successive approximations of the kind $C_{2n,1}^n$ with $n=1,2,3$ for $D_0=\gamma=1$. As can be seen, already $C_{2,1}^1$ captures the trend.

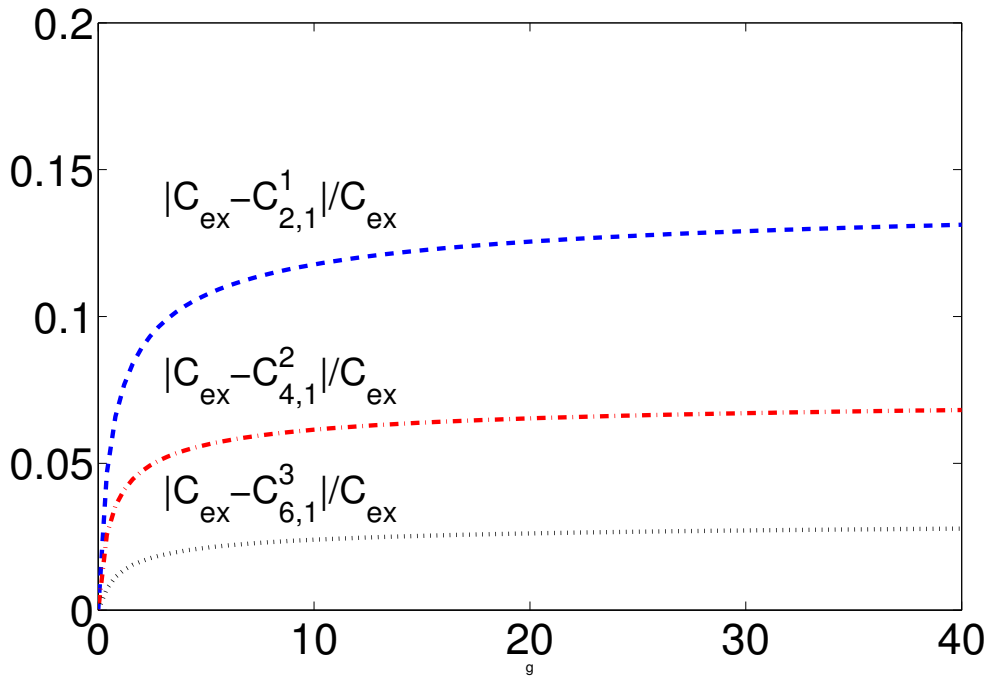

Fig. 3 – The relative deviation $(\langle\phi^2\rangle - C_{2n,1}^n)/\langle\phi^2\rangle$ for $D_0=\gamma=1$ with $n=1,2,3$. Note that the relative error saturates for $g\to\infty$, and that it decreases monotonically with $n$.

An interesting aspect of this convergence is that it is a bit unusual: we extract $\Gamma_{2n,1}(g)$ from a 1$^{st}$ order expansion of a high order moment, and plug it into a large order expansion of the two-point function. A more conventional convergence would be rather for $C_{2,n}^n$, which also happens here. However, the advantage of $C_{2n,1}^n$ over $C_{2,n}^n$ is that we know exactly $\Gamma_{2n,1}(g)$ (21) which requires solving only a quadratic polynomial equation, while for $\Gamma_{2,n}(g)$ we need to solve an n$^{th}$ order polynomial equation, which is in general not possible analytically. Furthermore, having field theories in mind, it is clear that $\Gamma_{2n,1}(g)$ is orders of magnitude simpler than $\Gamma_{2,n}(g)$, in terms of number of diagrams/terms needed.

**B. The SCE for the Kardar-Parisi-Zhang equation**

After showing the general ideology, structure as well as surprising convergence properties of the Self-Consistent Expansion (SCE) applied to the 0D example, we can now apply it to a stochastic field theory. We choose here the Kardar-Parisi-Zhang equation [4] given by

$$\frac{\partial \phi(\vec{x},t)}{\partial t} = \nu \nabla^2 \phi(\vec{x},t) + \frac{\lambda}{2}(\nabla \phi)^2 + \eta(\vec{x},t), \qquad (24)$$

which describes the fluctuation of an interface growing under ballistic deposition. The height function above a d-dimensional substrate is given by $\phi(\vec{x},t)$, and $\eta(\vec{x},t)$ is a noise-term modeling the fluctuation of the rate of deposition, which has a zero mean and is characterized by its second moment

$$\langle \eta(\vec{x},t)\eta(\vec{x}',t') \rangle = 2D_0 \delta^d(\vec{x}-\vec{x}')\delta(t-t'). \qquad (25)$$

This equation actually appears in many contexts in statistical physics [6] such as fluid flow, directed polymers in random media and more. Solutions of stochastic growth models such as (24) exhibiting scaling behavior are described by the time-dependent correlation function

$$\left\langle \left[\phi(\vec{x},t)-\phi(\vec{x}',t')\right]^2 \right\rangle = |\vec{x}-\vec{x}'|^{2\alpha} f\left(\frac{|\vec{x}-\vec{x}'|}{|t-t'|^z}\right), \qquad (26)$$

where $\alpha$ is the roughness exponent of the interface, $z$ is the dynamic exponent, and $f(u)$ is a scaling function.

Three difficulties appear when trying to move from the previous 0D example to a field theory like the KPZ equation. First, there is no Hamiltonian from which this equation can be derived, and therefore even the existence of a steady-state is not obvious, let alone its explicit form. Second, we need two basic quantities to describe growing surface. These are the steady-state (assuming that it exists) structure factor (also called the two-point function)

$$\Phi_q^{(2)} \equiv \left\langle \phi_q \phi_{-q} \right\rangle_S \qquad (27)$$

and its corresponding steady-state decay rate $\omega_q$, which describes the rate of decay of a disturbance of wave vector $q$ in steady state (the generalization of the friction coefficient above). A possible and widely used definition for $\omega_q$ is [1]

$$\omega_q^{-1} \equiv \frac{\int_0^\infty \left\langle \phi_{-q}(0)\phi_q(t) \right\rangle dt}{\left\langle \phi_q \phi_{-q} \right\rangle_S}, \qquad (28)$$

where $\left\langle \phi_{-q}(0)\phi_q(t) \right\rangle$ is a steady state average. Namely, $\phi_{-q}$ is measured in the steady state, the system is allowed to evolve freely for time $t$ and then $\phi_q$ is measured. The average of the product of those two measurements is then taken. From the scaling form (26), it follows that for small q's, $\Phi_q^{(2)}$ and $\omega_q$ behave as power laws in q, namely,

$$\Phi_q^{(2)} = Aq^{-\Gamma}. \qquad (29)$$

$$\omega_q = Bq^z. \qquad (30)$$

where $z$ is the dynamic exponent, and $\Gamma$ is related to the roughness exponent by $\alpha = (\Gamma - d)/2$.

In order to implement the SCE approach to KPZ we first take the Fourier transform of eqs. (24)-(25)

$$\frac{\partial \phi_q}{\partial t}(t) = -\nu_q \phi_q - \sum_{\ell,m} M_{q\ell m}\phi_\ell \phi_m + \eta_q(t), \qquad (31)$$

where $\nu_q = \nu q^2$ and $M_{q\ell m} = \frac{\lambda}{2\sqrt{\Omega}} \vec{\ell}\cdot\vec{m}\,\delta_{q,\ell+m}$, where $\Omega$ is the volume of the d-dimensional substrate and the noise correlations are now given by

$$\langle \eta_q(t)\eta_{q'}(t')\rangle = 2D_0 \delta_{q,-q}\delta(t-t'). \qquad (32)$$

In analogy with the 0D example above eq. (3), we rewrite this system in a Fokker-Planck form for the probability distribution functional $P(\{h_q\},t)$

$$\frac{\partial P}{\partial t} - \sum_q \frac{\partial}{\partial \phi_q}\left[D_0\frac{\partial}{\partial \phi_{-q}} + \nu_q \phi_q + \sum_{\ell,m} M_{q\ell m}\phi_\ell \phi_m\right]P = 0. \qquad (33)$$

As explained above, we now need to introduce two unspecified functions, denoted here $\hat{D}_q$ and $\hat{\omega}_q$, so that we get the following SCE scheme

$$\frac{\partial P}{\partial t} - \sum_q \frac{\partial}{\partial \phi_q}\left[\hat{D}_q \frac{\partial}{\partial \phi_{-q}} + \hat{\omega}_q\phi_q + \sum_{\ell,m} M_{q\ell m}\phi_\ell\phi_m + \Delta D\frac{\partial}{\partial \phi_{-q}} + \Delta\omega\phi_q\right]P = 0, \qquad (34)$$

with $\Delta D = D_0 - \hat{D}_q$, and $\Delta\omega = \nu_q - \hat{\omega}_q$. The two first terms in the sum, so far unspecified, are considered $0^{th}$ order, the $M_{q\ell m}$ term as $1^{st}$ order and last two difference terms are considered $2^{nd}$ order. (The exponents obtained in the end do net depend on that separation to $1^{st}$ and $2^{nd}$ order. The same exponents are obtained when all that is not zero order is taken as $1^{st}$ order. The separation is technically useful but no more than that.)

We consider next the steady state equation and multiply it by some functional $\mathbb{F}$ of the $\phi$'s to obtain a functional differential equation, which is the analog of equation (11) for the zero-dimensional case

$$D_0 \sum_p \left[\left\langle \frac{\partial^2 \mathbb{F}}{\partial \phi_p \partial \phi_{-p}}\right\rangle - \nu_p \left\langle \phi_p \frac{\partial \mathbb{F}}{\partial \phi_p}\right\rangle - \sum_{l,m} M_{plm}\left\langle \phi_l\phi_m \frac{\partial \mathbb{F}}{\partial \phi_p}\right\rangle\right] = 0 \qquad (35)$$

where all the averages are steady state averages.

In analogy with equation (13), we write

$$\sum_{\mathbf{p}} \hat{D}_{\mathbf{p}} \left\langle \frac{\partial^2 \mathbb{F}}{\partial \phi_{\mathbf{p}} \partial \phi_{-\mathbf{p}}} \right\rangle^{(n)} - \hat{\omega}_{\mathbf{p}} \left\langle \phi_{\mathbf{p}} \frac{\partial \mathbb{F}}{\partial \phi_{\mathbf{p}}} \right\rangle^{(n)}$$
$$- \sum_{\mathbf{l},\mathbf{m}} M_{\mathbf{plm}} \left\langle \phi_{\mathbf{l}} \phi_{\mathbf{m}} \frac{\partial \mathbb{F}}{\partial \phi_{\mathbf{p}}} \right\rangle^{(n-1)} - \Delta D \left\langle \frac{\partial^2 \mathbb{F}}{\partial \phi_{\mathbf{p}} \partial \phi_{=\mathbf{p}}} \right\rangle^{(n-2)} - \Delta \omega \left\langle \phi_{\mathbf{p}} \frac{\partial \mathbb{F}}{\partial \phi_{\mathbf{p}}} \right\rangle^{(n-2)} = 0 \quad (36)$$

Repeating the zero-dimensional process we start with $\mathbb{F} = \tfrac{1}{2}\phi_{\mathbf{q}}\phi_{-\mathbf{q}}$ and obtain

$$\left\langle \phi_{\mathbf{q}} \phi_{-\mathbf{q}} \right\rangle^{(0)} = \frac{\hat{D}_{\mathbf{q}}}{\hat{\omega}_{\mathbf{q}}} \equiv \Gamma_{\mathbf{q}} \quad (37)$$

This is now a basis for an expansion which gives, say, for second order

$$\left\langle \phi_{\mathbf{q}} \phi_{-\mathbf{q}} \right\rangle^{(2)} = \Gamma_{\mathbf{q}} + C_{\mathbf{q}}\{\Gamma_{\ell}, \omega_{\ell}\} \quad (38)$$

Requiring next that the second order result will not differ from the zero order result we are left with an equation, in the present case an integral equation, for the two functions $\Gamma$ and $\hat{\omega}$,

$$C_{\mathbf{q}}\{\Gamma_{\ell}, \hat{\omega}_{\ell}\} = 0. \quad (39)$$

We could get the same type of equation from any order of the expansion and how to construct an appropriate diagrammatic expansion to obtain $C_{\mathbf{q}}$ in a higher order expansion is explained in [23]. In any case this is just one integral equation and we need to fix two unknown functions. Now, considering the steady state is not enough and time dependence is required. The full detail may be found in [14] and will not be repeated here. The main idea now is to obtain a perturbation expansion for $\omega_{\mathbf{q}}$ (eq. (28)). The zero order equation is

$$\omega_{\mathbf{q}}^{(0)} = \hat{\omega}_{\mathbf{q}} \quad (40)$$

The second order expression looks like

$$\omega_{\mathbf{q}}^{(2)} = \hat{\omega}_{\mathbf{q}} + E_{\mathbf{q}}\{\Gamma_{\ell}, \hat{\omega}_{\ell}\} \quad (41)$$

Again we require that the perturbation does nothing and this results in a second integral equation.

$$E_{\mathbf{q}}\{\Gamma_{\ell}, \hat{\omega}_{\ell}\} = 0 \quad (42)$$

The two integral equations have now to be solved to yield the functions $\Gamma_q$ and $\hat{\omega}_q$. Since we are interested only in the exponents that give the small $q$ behavior of both functions, it turns out that the formidable task of solving two coupled non-linear integral equations can be avoided and depending on the problem, the exponents can be obtained either from simple power counting or in more interesting cases like KPZ from a solution of a transcendental equation.

**C. SCE for the time dependent structure factor**

Once we have the exponents, we can obtain the full time dependent structure factor, $\langle \phi_{-\mathbf{q}}(0)\phi_{\mathbf{q}}(t) \rangle$, using similar ideas. This time, however, it is more convenient to use the Langevin rather than the Fokker Planck formulation. First we take the Fourier transform with respect to time of equation (31),

$$i\omega\phi(\mathbf{q},\omega) + \nu_q \phi(\mathbf{q},\omega) + \sum N_{\mathbf{qlm}}\delta_{\sigma+\tau,\omega}\phi(\mathbf{l},\sigma)\phi(\mathbf{m},\tau) = \eta(\mathbf{q},\omega) \qquad (43)$$

where $N_{\mathbf{qlm}} = M_{\mathbf{qlm}}/\sqrt{T}$, $T$ being an assumed periodicity in time to be taken eventually to infinity and the noise correlations are given by $\langle \eta(\mathbf{q},\omega)\eta(-\mathbf{q},-\omega) \rangle = 2D_0$. Equation (43) is expressed now in the form

$$\left[ \left(i\omega + \hat{\omega}_q\right)\phi(\mathbf{q},\omega) - \eta^{(0)}(\mathbf{q},\omega) \right] + \left[ \sum_{\mathbf{l},\mathbf{m}} N_{\mathbf{qlm}}\delta_{\sigma+\tau,\omega}\phi(\mathbf{l},\sigma)\phi(\mathbf{m},\tau) - \eta^{(1)}(\mathbf{q},\omega) \right]$$
$$+ \left[ \nu_q - \hat{\omega}_q \right]\phi(\mathbf{q},\omega) = 0, \qquad (44)$$

where the three terms on the left hand side of the above are taken as zero first and second order from left to right. The original noise has been broken into two uncorrelated noises such that

$$\langle \eta^{(0)}(\mathbf{q},\omega)\eta^{(0)}(-\mathbf{q},-\omega) \rangle = 2\hat{D}(\mathbf{q},\omega) \qquad (45)$$

The zero order solution is given by

$$\phi(\mathbf{q},\omega) = \frac{\eta^{(0)}(\mathbf{q},\omega)}{i\omega + \hat{\omega}_q} \qquad (46)$$

which is a starting point for a perturbation expansion that gives $\phi(\mathbf{q},\omega)$ to a required order. Next $\phi(\mathbf{q},\omega)$ is multiplied into its complex conjugate, average over the noise is taken and only terms to the order of the expansion of $\phi$ are kept. This results in an expansion for $\Phi(q,\omega)$, which is the Fourier transform of the time dependent structure factor. The zero order result is

$$\Phi^{(0)}(q,\omega) = \frac{4\hat{D}(q,\omega)}{\omega^2 + \hat{\omega}_q^2}, \tag{47}$$

The expansion, say to second order has now the form

$$\Phi^{(2)}(q,\omega) = \Phi^{(0)}(q,\omega) + L_{q,\omega}\{\Phi^{(0)}(l,\sigma)\} \tag{48}$$

The same requirement as before, that the perturbation does not affect the result, leaves us with the integral equation,

$$L_{q,\omega}\{\Phi^{(0)}(l,\sigma)\} = 0. \tag{49}$$

Recall that now $\Phi^{(0)}$ is the result that we are looking for. Again, unrelated to the SCE itself, asymptotic behavior of the time dependent structure factor does not need the full solution of the integral equation [20], [23].

We hope that the present article will tempt the reader to spend the necessary time to study the details of the method, which will result in its application to more problems of interest or to modifications and improvement of the method itself.

**Acknowledgement**